\documentclass[english,pop,showpacs,preprint,superscriptaddress]{revtex4-1}
\usepackage{lmodern}

\usepackage[T1]{fontenc}
\usepackage[utf8]{inputenc}
\usepackage{amsmath}
\usepackage{amssymb}
\usepackage{graphicx}
\usepackage{esint}

\makeatletter
\@ifundefined{date}{}{\date{}}

\usepackage{babel}

\usepackage{babel}

\usepackage{babel}

\usepackage{babel}

\usepackage{babel}

\usepackage{babel}

\usepackage{babel}

\makeatother

\usepackage{babel}
\begin{document}

\title{On the correspondence between classical geometric phase of gyro-motion
and quantum Berry phase}

\author{Hongxuan Zhu}

\affiliation{Plasma Physics Laboratory, Princeton University, Princeton, NJ 08543}

\author{Hong Qin}

\affiliation{Plasma Physics Laboratory, Princeton University, Princeton, NJ 08543}

\affiliation{School of Nuclear Science and Technology and Department of Modern
Physics, University of Science and Technology of China, Hefei, Anhui
230026, China}
\begin{abstract}
We show that the geometric phase of the gyro-motion of a classical
charged particle in a uniform time-dependent magnetic field described
by Newton's equation can be derived from a coherent Berry phase for
the coherent states of the Schrödinger equation or the Dirac equation.
This correspondence is established by constructing coherent states
for a particle using the energy eigenstates on the Landau levels and
proving that the coherent states can maintain their status of coherent
states during the slow varying of the magnetic field. It is discovered
that orbital Berry phases of the eigenstates interfere coherently
to produce an observable effect (which we termed ``coherent Berry
phase''), which is exactly the geometric phase of the classical gyro-motion.
This technique works for particles with and without spin. For particles
with spin, on each of the eigenstates that makes up the coherent states,
the Berry phase consists of two parts that can be identified as those
due to the orbital and the spin motion. It is the orbital Berry phases
that interfere coherently to produce a coherent Berry phase corresponding
to the classical geometric phase of the gyro-motion. The spin Berry
phases of the eigenstates, on the other hand, remain to be quantum
phase factors for the coherent states and have no classical counterpart. 
\end{abstract}
\maketitle

\section{Introduction}

Berry phase \cite{Berry1984} of a quantum system is an important
physical effect that has been discussed in depth \cite{Berry1985,Simon1983,AA1987}.
Because Berry phase, as a quantum phase in the wave function, depends
only on the geometric path of the system, it is also called geometric
phase. The phenomena of geometric phase also exist in classical systems,
known as the Hannay angle \cite{Hannay1985}. Geometric phase also
exists in many systems in plasma physics \cite{Bhattacharjee1992,Liu11,Liu12,Burby13,Lan16}.
To avoid confusion, geometric phase is used only for classical systems
in this paper.

In a magnetized plasma, charged particles gyrate in the plane perpendicular
to the magnetic field, exerting helical orbits. This gyro-motion of
a charged particle can be characterized by a dynamic gyro-phase around
the magnetic field. In a strongly magnetized plasma, the fast gyro-motion
of charged particles leads to the temporal and spatial scale separation,
and is usually averaged out in the magneto-hydrodynamic and traditional
gyro-kinetic theories. However, the gyro-phase itself still carries
important information and plays an important role in modern gyro-kinetic
theories \cite{Qin00,Qin2007,Yu09,Burby12}. Recently, Liu and Qin
\cite{Liu11} discussed the gyro-motion of a charged particle in a
spatially uniform, time-dependent magnetic field \cite{Qin06}. It
was found that when the magnetic field returns to its original direction,
apart from the phase advance produced by the gyro-motion, there is
an additional geometric phase in the gyro-phase, which equals to the
solid angle $\Omega$ spanned by the trace of the magnetic field unit
vector $\boldsymbol{b}$ on the unit sphere $S^{2}$. On the other
hand, it is well known that the Berry phase associated with an electron
spin eigenstate under the same change of the magnetic field is $\pm\frac{1}{2}\Omega$
\cite{Berry1984}, whose sign depends on the spin direction. Ref.\,\cite{Liu11}
discussed the similarities and differences between the geometric phase
in a charged particle's gyro-motion and the Berry phase for the electron
spin in quantum mechanics. However, no direct connection was found
in their paper. Even though the gyro-motion is not the classical counterpart
of the quantum spin, the similarities in these two geometric phases
may still imply certain connections in a deeper level.

In this paper, we show a direct correspondence between the classical
geometric phase of the gyro-motion and the Berry phase of the underlying
quantum system, based on the use of coherent states from the Schrödinger
equation. The Berry phases come from the orbital angular momentum
eigenstates on the Landau levels \cite{Landau1930}, rather than the
spin eigenstates. The Berry phase is governed by the Schrödinger equation,
while the geometric phase of the gyro-motion is governed by Newton's
equation. The connection between the two reveals the identical physical
and geometric nature for the two phases. Historically, the problem
being studied here is about the quantum-classical correspondence.
Berry studied the correspondence between Berry phase and Hannay angle
from the semi-classical point of view \cite{Berry1985}, while the
idea of using coherent state to establish such correspondence in a
1-dimensional non-degenerate system was first developed by Maamache
et al \cite{Maamache90}. Here we study the correspondence problem
for a 2-dimensional degenerate system in the context of charged particle
dynamics in an external magnetic field.

The correspondence is establish through three steps. First, we construct
a coherent state to represent the gyro-motion of a charged particle
in a uniform time-independent magnetic field. Then we calculate the
Berry phase for each component that makes up the coherent state during
the slowly varying of the magnetic field. Lastly, we prove that the
interference of these components after gaining their Berry phases
results in an additional phase that naturally enters the complex variable
that is used to define the coherent state. We termed this additional
phase ``coherent Berry phase'', which turns out to be exactly the
classical geometric phase of the gyro-motion. To further clarify the
relationship between the geometric phase of the gyro-motion and the
Berry phase of a charged particle with spin, we will also analyze
electrons with spin governed by the Dirac equation and show that the
Berry phase of an eigenstate in the non-relativistic limit consists
of two parts, the orbital part and the spin part. The orbital Berry
phases of the eigenstates interfere coherently to produce a coherent
Berry phase corresponding to the classical geometric phase of gyro-motion,
while the spin Berry phases have no classical counterparts, as expected.

The term ``coherent Berry phase'' in this paper has not been used
before. It is not an overall phase factor multiplying the wave function,
but the argument $\gamma_{C}$ that enters the complex variable $w_{0}(T)$
which is used to define a coherent state. It is an observable effect
resulting from the coherent interference of all the eigenstates that
constitute the coherent state, each of them having gained a Berry
phase $\gamma_{n}=n\gamma_{C}$ (See Sec.~IV for detailed discussions).
We want to distinguish it from the concept of ``Berry phase of a
coherent state'', which is indeed an overall phase factor in front
of the coherent state, and was calculated, for example, in Ref.~\cite{Yang07,Chaturvedi87}. 

The paper is organized as follows. In Sec.\,II we briefly review
the derivation of the geometric phase of a charged particle's gyro-motion.
We review in Sec.\,III the derivation of the Landau levels and construct
coherent states for a charged particle in a uniform time-independent
magnetic field. The Berry phase associated with a coherent state is
calculated and the correspondence between the geometric phase of the
gyro-motion is established in Sec.\,IV. In Sec.\,V, we calculate
the Berry phases of an electron described by the Dirac equation and
analyze the Berry phases due to orbital and spin degrees of freedom.

\section{Classical geometric phase of a charged particle's gyro-motion}

In this section, we review the derivation \cite{Liu11} of the classical
geometric phase of the gyro-motion for a classical charged particle
in a time-dependent magnetic field. Consider a classical charged particle
with charge $q$ and mass $\mu$ in a spatially-uniform but time-dependent
magnetic field $\boldsymbol{B}=B(t)\boldsymbol{b}(t)$. Newton's equation
for the particle is 
\begin{equation}
\frac{d\boldsymbol{v}}{dt}=\omega_{d}\boldsymbol{v}\times\boldsymbol{b},
\end{equation}
where $\omega_{d}(t)=qB(t)/\mu$ is the gyro-frequency. To define
the gyro-phase, we need to select a frame. Choose two unit vectors
$\boldsymbol{e}_{1}$ and $\boldsymbol{e}_{2}$ perpendicular to $\boldsymbol{b}$
for every possible $\boldsymbol{b}$ such that $\boldsymbol{e}_{1}\cdot\boldsymbol{e}_{2}=0$
and $\boldsymbol{e}_{1}\times\boldsymbol{e}_{2}=\boldsymbol{b}$.
Note that there is a freedom in choosing $(\boldsymbol{e}_{1},\boldsymbol{e}_{2})$.
Particle velocity can be decomposed in the frame $(\boldsymbol{e}_{1},\boldsymbol{e}_{2},\boldsymbol{b})$
as

\begin{equation}
\boldsymbol{v}=v_{\parallel}\boldsymbol{b}+v_{\perp}\cos\theta\boldsymbol{e}_{1}+v_{\perp}\sin\theta\boldsymbol{e}_{2},
\end{equation}
where $\theta$ is the gyro-phase. Following Ref.\,\cite{Liu11},
the dynamic equation for $\theta$ is 
\begin{gather}
\frac{d\theta}{dt}=-\left[\omega_{d}(t)+\omega_{g}(t)+\omega_{a}(t)\right],\label{eq:dthe}\\
\omega_{g}(t)=\frac{d\boldsymbol{e}_{1}}{dt}\cdot\boldsymbol{e}_{2},\\
\omega_{a}(t)=\frac{v_{\parallel}}{v_{\perp}}\frac{d\boldsymbol{b}}{dt}\cdot(\cos\theta\boldsymbol{e}_{2}-\sin\theta\boldsymbol{e}_{1}),
\end{gather}
where $\omega_{d}(t)$ is the dynamic contribution due to gyro-motion,
$\omega_{g}(t)$ is the geometric contribution, and $\omega_{a}(t)$
is the adiabatic contribution for reasons soon to be clear. The negative
sign on the right-hand side of Eq.\,(\ref{eq:dthe}) is due the choice
of coordinate. Liu and Qin \cite{Liu11} proved that if the magnetic
field changes slowly, i.e., 
\begin{flalign}
 & |\frac{1}{\omega_{d}B}\frac{d\boldsymbol{B}}{dt}|\sim\epsilon\ll1,\nonumber \\
 & |\frac{1}{\omega_{d}^{2}B}\frac{d^{2}\boldsymbol{B}}{dt^{2}}|\sim\epsilon^{2}\ll1,\label{eq:adiabatic}
\end{flalign}
then the phase advances due to the dynamic, the geometric, and the
adiabatic phase satisfy the following ordering, 
\begin{gather*}
\Delta\theta_{d}:\Delta\theta_{g}:\Delta\theta_{a}\sim1:\epsilon:\epsilon^{2},\\
\Delta\theta_{d}\equiv-\int_{0}^{T}\omega_{d}dt,\\
\Delta\theta_{g}\equiv-\int_{0}^{T}\omega_{d}dt,\\
\Delta\theta_{a}\equiv-\int_{0}^{T}\omega_{a}dt.
\end{gather*}
For a slowing evolving system, the leading order correction to the
dynamic phase $\Delta\theta_{d}$ is the geometric phase $\Delta\theta_{g}$.
Assume that the system starts evolving from $t=0$, and at $t=T$
the magnetic field returns to its original position, i.e., $\boldsymbol{b}(T)=\boldsymbol{b}(0)$.
The fact that the frame $(\boldsymbol{e}_{1},\boldsymbol{e}_{2},\boldsymbol{b})$
is defined in a single-valued manner implies that $\boldsymbol{e}_{1}(T)=\boldsymbol{e}_{1}(0)$
and $\boldsymbol{e}_{2}(T)=\boldsymbol{e}_{2}(0)$. The trace of $\boldsymbol{b}(t)$
during the time forms a closed loop $C$ on $S^{2}$. The geometric
phase is calculated to be 
\begin{equation}
\Delta\theta_{g}=-\int_{0}^{T}\omega_{g}dt=-\oint_{C}d\boldsymbol{e}_{1}\cdot\boldsymbol{e}_{2}.\label{eq:solidangle}
\end{equation}
The last integration is along the closed loop $C$ on $S^{2}$. In
the spherical coordinates $(\zeta,\phi)$ with $\boldsymbol{b}=(\sin\zeta\cos\phi,\sin\zeta\sin\phi,\cos\zeta)$,
we can choose $\boldsymbol{e}_{1}=(\cos\zeta\cos\phi,\cos\zeta\sin\phi,-\sin\zeta)$
and $\boldsymbol{e}_{2}=(-\sin\phi,\cos\phi,0)$. The geometric phase
becomes 
\begin{equation}
\Delta\theta_{g}=-\oint_{C}\cos\zeta d\phi=-\Omega,
\end{equation}
where $\Omega$ is the solid angle expanded by $C$. We note that
$\Omega$ does not depend on the choice of frame. For a different
frame $(\boldsymbol{e}_{1}',\boldsymbol{e}_{2}',\boldsymbol{b})$
specified by a coordinate transformation 
\begin{equation}
\begin{cases}
\boldsymbol{e}_{1}' & =\cos\psi\boldsymbol{e}_{1}+\sin\psi\boldsymbol{e}_{2},\\
\boldsymbol{e}_{2}' & =-\sin\psi\boldsymbol{e}_{1}+\cos\psi\boldsymbol{e}_{2},
\end{cases}
\end{equation}
we have $d\boldsymbol{e}_{1}'\cdot\boldsymbol{e}_{2}'=d\boldsymbol{e}_{1}\cdot\boldsymbol{e}_{2}+d\psi$.
For $\boldsymbol{b}(T)=\boldsymbol{b}(0)$, $\psi(T)=\psi(0)$ and
thus $\oint d\psi=0$. Therefore, the geometric phase is unique when
the magnetic field returns to its original direction.

\section{Landau levels and coherent states for spinless particles }

In this section, we construct coherent states of a spinless charged
particle in a uniform magnetic field described by the Schrödinger
equation. The energy eigenstates are infinitely degenerate in each
of the energy levels which are known as Landau levels \cite{Landau1930}.
From these eigenstates, we can construct a coherent state which is
a non-diffusive wave packet gyrating around the magnetic field and
corresponds to a classical charged particle. Several authors have
discussed how to construct coherent states \cite{Malkin1969,Feldman1970,Kowalsk2005,Murayama,Shankar94}.
Here we review these results using the notation of Refs.\,\cite{Murayama,Shankar94}.
It's assumed that the particle has positive charge $q>0$ . For negative
charge, the definitions will be modified accordingly, as will be seen
in Sec.\,V.

The Hamiltonian for the charged particle of charge $q$ and mass $\mu$
in a uniform magnetic field $\boldsymbol{B}=B_{0}\boldsymbol{e}_{z}$
is 
\begin{equation}
H=\frac{(\boldsymbol{P}-q\boldsymbol{A})^{2}}{2\mu},
\end{equation}
where $\boldsymbol{P}=-i\hbar\nabla$ is the canonical momentum operator,
and $\boldsymbol{A}$ is the magnetic vector potential satisfying
$\nabla\times\boldsymbol{A}=\boldsymbol{B}.$ The kinetic momentum
operator is $\boldsymbol{\ensuremath{\pi}}=\boldsymbol{P}-q\boldsymbol{A}$,
whose $x$ and $y$-components satisfy the commutation relation 
\begin{equation}
[\pi_{x},\pi_{y}]=i\hbar q(\partial_{x}A_{y}-\partial_{y}A_{x})=i\hbar qB_{0}.
\end{equation}
We can then define creating and annihilating operators $a^{\dagger}$
and $a$ as 
\begin{equation}
\begin{cases}
a^{\dagger}=\sqrt{\frac{1}{2\hbar qB_{0}}}(\pi_{x}-i\pi_{y}),\\
a=\sqrt{\frac{1}{2\hbar qB_{0}}}(\pi_{x}+i\pi_{y}),
\end{cases}
\end{equation}
and prove the commutation relation $[a,a^{\dagger}]=1$.

The Hamiltonian can be written as 
\begin{equation}
H=\hbar\omega_{d}(a^{\dagger}a+\frac{1}{2}),
\end{equation}
where $\omega_{d}=qB/\mu$. Since a particle moves freely along the
magnetic field, the parallel motion $\boldsymbol{P}_{z}^{2}/2\mu$
is not included for the moment (the discussion on $\boldsymbol{P}_{z}$
can be found in the Appendix). The Hamiltonian is in the same form
as that of a 1D simple harmonic oscillator. Choosing $\boldsymbol{A}$
to be the rotationally symmetric form $\boldsymbol{A}=(-\frac{1}{2}B_{0}y,\frac{1}{2}B_{0}x,0)$,
and using complex variables $w\equiv x+iy,$ we express the creating
and annihilating operators as 
\begin{equation}
\begin{cases}
a^{\dagger}=-i\sqrt{\frac{\hbar}{2qB_{0}}}(2\partial_{w}-\frac{qB_{0}}{2\hbar}\bar{w}),\\
a=-i\sqrt{\frac{\hbar}{2qB_{0}}}(2\partial_{\bar{w}}+\frac{qB_{0}}{2\hbar}w).
\end{cases}
\end{equation}
Here, $\partial_{w}=\frac{1}{2}(\partial_{x}-i\partial_{y})$, $\partial_{\overline{w}}=\frac{1}{2}(\partial_{x}+i\partial_{y})$,
and $w$ and $\overline{w}$ are treated as independent variables
(the over-bar means complex conjugate). The ground state $\psi(w,\bar{w})$
is obtained by solving $a\psi=0$, i.e., 
\begin{equation}
-i\sqrt{\frac{\hbar}{2qB_{0}}}(2\partial_{\bar{w}}+\frac{qB_{0}}{2\hbar}w)\psi(w,\bar{w})=0.
\end{equation}
The solution is $\psi(w,\bar{w})=g(w)e^{-qB_{0}w\bar{w}/4\hbar}$,
where $g(w)$ is an arbitrary analytical function. The arbitrariness
of $g(w)$ indicates the infinite degeneracy of the ground states.
With the choice of $g(w)=w^{m},\,m=0,1,2,...$, a set of ground states
can be obtained, 
\begin{gather}
\psi_{0,m}=N_{m}w^{m}e^{-qB_{0}w\bar{w}/4\hbar},\label{eq:groundstates}\\
N_{m}=\left[\pi m!\left(\frac{2\hbar}{qB_{0}}\right)^{m+1}\right]^{-\frac{1}{2}},
\end{gather}
where $N_{m}$ is the normalization factor. Excited states are obtained
using the creating operator, 
\begin{equation}
\psi_{n,m}=\frac{(a^{\dagger})^{n}}{\sqrt{n!}}\psi_{0,m}.
\end{equation}
It is easy to verify that $H\psi_{n,m}=\hbar\omega_{d}(n+\frac{1}{2})$.
The eigenstates $\psi_{n,m}$ covers all the Landau levels, with each
$n$ representing an energy level $E_{n}=\hbar\omega_{d}(n+\frac{1}{2})$
with infinite degeneracy. They are all the eigenstates of the angular
momentum operator $\boldsymbol{L}_{z}=-i\hbar\partial_{\theta}$.
In the polar coordinates with $w=\rho e^{i\theta}$, it is straightforward
to show that $\psi_{n.m}\propto e^{-i(n-m)\theta}$ and that they
are orthogonal to each other, i.e., $\left<\psi_{n,m}|\psi_{n',m'}\right>=0$,
for any $(n',m')\neq(n,m)$.

Since the motion perpendicular to the magnetic field is essentially
4-dimensional in phase space ($x,y,\pi_{x},\pi_{y}$), one pair of
creating and annihilating operators $(a^{\dagger},a)$ is incomplete.
Indeed, there exists another pair of creating and annihilating operators
($b^{\dagger},b)$, defined as
\begin{equation}
\begin{cases}
b^{\dagger}=\sqrt{\frac{qB_{0}}{2\hbar}}(X-iY),\\
b=\sqrt{\frac{qB_{0}}{2\hbar}}(X+iY),
\end{cases}
\end{equation}
where $(X,Y)=(x+\pi_{y}/\mu\omega_{d},y-\pi_{x}/\mu\omega_{d})$ is
the guiding center position operator. Using the complex variable $w\equiv x+iy$,
we have
\begin{equation}
\begin{cases}
b^{\dagger}=\sqrt{\frac{\hbar}{2qB_{0}}}(-2\partial_{\bar{w}}+\frac{qB_{0}}{2\hbar}w),\\
b=\sqrt{\frac{\hbar}{2qB_{0}}}(2\partial_{w}+\frac{qB_{0}}{2\hbar}\bar{w}).
\end{cases}
\end{equation}
It can be verified that $[b,a]=[b,a^{\dagger}]=[b^{\dagger},a]=[b^{\dagger},a^{\dagger}]=0$,
thus $[b,H]=[b^{\dagger},H]=0$. Also, the degenerate eigenstates
on each Landau level is related by $b^{\dagger}$ and $b$,
\begin{equation}
\psi_{n,m}=\frac{(b^{\dagger})^{m}}{\sqrt{m!}}\psi_{n,0}.
\end{equation}
Thus two pairs of creating and annihilating operators $(a^{\dagger},a)$
and $(b^{\dagger},b)$ give the complete description of motion perpendicular
to the magnetic field.

\begin{figure}
\includegraphics[scale=0.5]{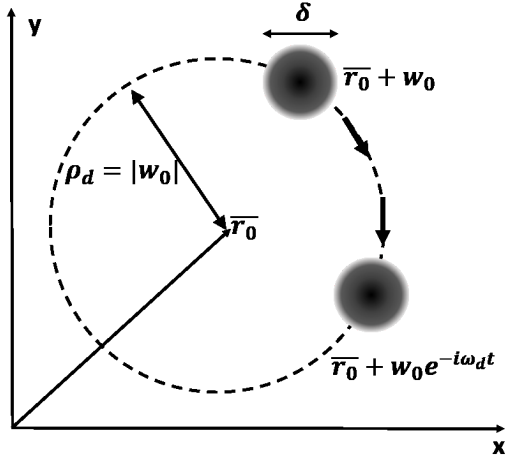}\caption{Illustration of the coherent state described by Eq.\,(\ref{eq:coherent}).\label{fig:Illustration}}
\end{figure}

A coherent state is constructed in a way similar to that of a simple
harmonic oscillator. Let $f=-i\sqrt{qB_{0}/2\hbar}w_{0}$ and $R=\sqrt{qB_{0}/2\hbar}r_{0}$
where $w_{0}$ and $r_{0}$ are complex variables, a coherent state
is 
\begin{flalign}
 & \Psi_{w_{0},r_{0}}=e^{-\frac{|f|^{2}+|R|^{2}}{2}}e^{fa^{\dagger}+Rb^{\dagger}}\psi_{0,0}\nonumber \\
 & =e^{-\frac{qB_{0}}{4\hbar}w_{0}\bar{w_{0}}}e^{-\frac{w_{0}}{2}(2\partial_{w}-\frac{qB_{0}}{2\hbar}\bar{w})}\left[e^{\frac{r_{0}}{2}(-2\partial_{\bar{w}}+\frac{qB_{0}}{2\hbar}w)}(N_{0}e^{-\frac{qB_{0}w\bar{w}}{4\hbar}})\right]\nonumber \\
 & =N_{0}e^{-\frac{qB_{0}}{4\hbar}(|w|^{2}+|w_{0}|^{2}+|r_{0}|^{2}+2r_{0}w_{0}-2r_{0}w-2w_{0}w)}.
\end{flalign}
The probability distribution of $\Psi_{w_{0},r_{0}}$ is 
\begin{equation}
|\Psi_{w_{0},r_{0}}|^{2}=|N_{0}|^{2}e^{-\frac{qB_{0}}{2\hbar}(w-\bar{r_{0}}-w_{0})(\bar{w}-r_{0}-\bar{w}_{0})},
\end{equation}
which describes a Gaussian wave packet in the $x$-$y$ plane. It
centers at $w\equiv x+iy=\bar{r_{0}}+w_{0}$ and has a characteristic
width $\delta=\sqrt{\hbar/qB_{0}}$. To obtain the time evolution
of $\Psi_{w_{0},r_{0}}$, we decompose it into eigenstates on the
Landau levels, 
\begin{flalign}
\Psi_{w_{0},r_{0}}=e^{-\frac{|f|^{2}+|R|^{2}}{2}}e^{fa^{\dagger}+Rb^{\dagger}}\psi_{0,0}=e^{-\frac{|f|^{2}+|R|^{2}}{2}}\sum_{n=0}^{+\infty}\sum_{m=0}^{+\infty}\frac{f^{n}}{\sqrt{n!}}\frac{R^{m}}{\sqrt{m!}}\psi_{n,m}.\label{eq:coherent0}
\end{flalign}
The coherent state evolves according to how each eigenstates evolves,
\begin{flalign}
\Psi_{w_{0},r_{0}}(t) & =e^{-\frac{|f|^{2}+|R|^{2}}{2}}\sum_{n=0}^{+\infty}\sum_{m=0}^{+\infty}\frac{f^{n}}{\sqrt{n!}}\frac{R^{m}}{\sqrt{m!}}\psi_{n,m}e^{iE_{n}(t)}.\nonumber \\
 & =e^{-\frac{|f|^{2}+|R|^{2}}{2}}\sum_{n=0}^{+\infty}\sum_{m=0}^{+\infty}\frac{f^{n}}{\sqrt{n!}}\frac{R^{m}}{\sqrt{m!}}\psi_{n,m}e^{-i\omega_{d}(n+\frac{1}{2})t}\nonumber \\
 & =e^{-\frac{i\omega_{d}t}{2}}e^{-\frac{|f(t)|^{2}+|R|^{2}}{2}}\sum_{n=0}^{+\infty}\sum_{m=0}^{+\infty}\frac{[f(t)]^{n}}{\sqrt{n!}}\frac{R^{m}}{\sqrt{m!}}\psi_{n,m}\nonumber \\
 & =e^{-\frac{i\omega_{d}t}{2}}\Psi_{w_{0}(t),r_{0}},\label{eq:coherent}
\end{flalign}
where $f(t)=-i\sqrt{qB_{0}/2\hbar}w_{0}(t)$ and $w_{0}(t)=w_{0}e^{-i\omega_{d}t}$.
We see that $\Psi_{w_{0},r_{0}}(t)$ still describes a Gaussian wave
packet, but its center has moved to $w=\bar{r_{0}}+w_{0}e^{-i\omega_{d}t}$.
The coherent state $\Psi_{w_{0},r_{0}}(t)$ does not diffuse, thus
represents the gyro-motion of a charged particle in a uniform time-independent
magnetic field, with guiding center at $\bar{r_{0}}$, gyro-frequency
$\omega_{d}$ and gyro-radius $\rho_{d}=|w_{0}|$. $w_{0}$ and $r_{0}$
are two complex variables, which have four degrees of freedom, thus
will give a complete description of all the coherent states. An illustration
of the coherent state described by Eq.\,(\ref{eq:coherent}) is shown
in Fig. \ref{fig:Illustration}.

Because the magnetic field is spatially homogeneous, the system has
translational invariance perpendicular to the magnetic field. Thus
we can perform a coordinate transformation to move the guiding center
to the origin. This coordinate transformation is accompanied by a
gauge transformation to make the vector potential $\boldsymbol{A}$
rotationally invariant around the new origin. Such a gauge transformation
is
\begin{align}
\boldsymbol{A} & \rightarrow\boldsymbol{A}-\nabla\chi,\\
\Psi_{w_{0},r_{0}} & \rightarrow e^{-\frac{iq}{\hbar c}\chi}\Psi_{w_{0},r_{0}},\\
\chi(x,y) & =\frac{B_{0}}{2}\left[Re(\bar{r_{0}})y-Im(\bar{r_{0}})x\right].
\end{align}
It's found that after the gauge transformation, the wave function
of the coherent state becomes
\begin{equation}
e^{-\frac{iq}{\hbar c}\chi}\Psi_{w_{0},r_{0}}=N_{0}e^{-\frac{qB_{0}}{4\hbar}[|w-\bar{r_{0}}|^{2}-2w_{0}(w-\bar{r_{0}})+|w_{0}|^{2}]}.
\end{equation}
Thus a coordinate transformation $w\rightarrow w-\bar{r_{0}}$ will
transform the wave function into
\begin{equation}
N_{0}e^{-\frac{qB_{0}}{4\hbar}[|w|^{2}-2w_{0}w+|w_{0}|^{2}]}=\Psi_{w_{0},0}.
\end{equation}
Hence, the time-evolution of the coherent state does not depend on
$r_{0}$ once we perform a gauge transformation and a coordinate transformation.
In the following discussion we choose $r_{0}=0$ for simplicity, and
the coherent state can be simplified to
\begin{equation}
\Psi_{w_{0}}\equiv\Psi_{w_{0},0}=e^{-\frac{|f|^{2}}{2}}\sum_{n=0}^{+\infty}\frac{f^{n}}{\sqrt{n!}}\psi_{n,0}.
\end{equation}

\section{The Berry phase associated with coherent states for spinless particles}

We show in this section that a coherent Berry phase can be naturally
defined for the coherent states when the magnetic field evolves slowly
with time, and this coherent Berry phase is exactly the geometric
phase for the classical gyro-motion. We assume for simplicity that
the magnitude of the magnetic field does not change, and only the
field direction changes, i.e., $\boldsymbol{B}(t)=B_{0}\boldsymbol{b}(t)$.
At $t=T$, the magnetic field returns to its original state, i.e.
$\boldsymbol{b}(T)=\boldsymbol{b}(0)=\boldsymbol{e}_{z}$. Then the
trajectory of $\boldsymbol{b}(t)$ on $S^{2}$ forms a closed loop
$C$. As in Sec.\,II, for each $\boldsymbol{b}(t)$, we choose unit
vectors $\boldsymbol{e}_{1}$ and $\boldsymbol{e}_{2}$ such that
$\boldsymbol{e}_{1}\cdot\boldsymbol{e}_{2}=0$ and $\boldsymbol{e}_{1}\times\boldsymbol{e}_{2}=\boldsymbol{b}$.
The Hamiltonian depends on $\boldsymbol{b}(t)$ parametrically. For
a given $\boldsymbol{b}$, 
\begin{align}
H[\boldsymbol{b}] & =\frac{\left(\boldsymbol{P}-q\boldsymbol{A}[\boldsymbol{b}]\right)^{2}}{2\mu},\\
\boldsymbol{A}[\boldsymbol{b}] & =-\frac{1}{2}B_{0}\boldsymbol{r}\times\boldsymbol{b},
\end{align}
where $\boldsymbol{r}=(x,y,z)$ is the coordinate vector in the Cartesian
frame of $\mathbb{R}^{3}$, and $\boldsymbol{P}=-i\hbar\nabla$. The
eigenstates of $H[\boldsymbol{b}]$ are $\psi_{n,m}[\boldsymbol{b}]=\psi_{n,m}\left(w[\boldsymbol{b}],\bar{w}[\boldsymbol{b}]\right)$,
where $w[\boldsymbol{b}]=\rho[\boldsymbol{b}]e^{i\theta[\boldsymbol{b}]}$,
$\bar{w}[\boldsymbol{b}]=\rho[\boldsymbol{b}]e^{-i\theta[\boldsymbol{b}]}$
and 
\begin{align}
\rho[\boldsymbol{b}] & =\sqrt{|\boldsymbol{r}|^{2}-[\boldsymbol{r}\cdot\boldsymbol{b}]},\label{eq:polarrho}\\
\theta[\boldsymbol{\boldsymbol{b}}] & =\arccos\frac{[\boldsymbol{r}-(\boldsymbol{r}\cdot\boldsymbol{b})\boldsymbol{b}]\cdot\boldsymbol{e}_{1}}{\rho[\boldsymbol{b}]}.\label{eq:polartheta}
\end{align}
In the above equations, the notation $[\boldsymbol{b}]$ denotes the
parametric dependence on $\boldsymbol{b}$. For example, $\psi_{n,m}[\boldsymbol{b}(t)]$
is an eigenstate corresponding to the $\boldsymbol{b}$ at the instant
of $t$. It is not the solution of the time-dependent Schrödinger
equation.

Now the question is, if the system is at a coherent state $\Psi_{w_{0}}$
at $t=0$, what is the state of the system at $t=T$ under a slow
evolution of $\boldsymbol{b}(t)$? To answer this question, we first
look at how each eigenstate evolves. According to the well-known adiabatic
theorem \cite{Messiah1962}, it is expected that each eigenstate $\psi_{n,0}(t=0)$
is evolved into the eigenstate $\psi_{n,0}[\boldsymbol{b}(t)]$. However,
the adiabatic theorem in its general form only applies to non-degenerate
systems \cite{AA1987,Messiah1962,Holstein1989}, which casts doubt
on this expectation. Fortunately, we can prove that the adiabatic
theorem still holds for eigenstates $\psi_{n,m}$ on the Landau levels.
Thus when the magnetic field changes slowly enough, each energy eigenstate
$\psi_{n,0}[\boldsymbol{b}(0)]$ at $t=0$ will always be the eigenstate
and independently gain a Berry phase . The proof is presented in the
Appendix.

However, there is still no guarantee that a coherent state at $t=0$
will remain to be a coherent state at $t>0,$ even though the adiabatic
theorem holds and each eigenstate that makes up the coherent state
maintains its eigenstate status. This is because the Berry phase of
each eigenstate may not be consistent with the requirement of the
coherent state. Fortunately again, we find that for the problem presently
investigated, each eigenstate gains a Berry phase in such a way that
the coherent state at $t=0$ maintains its status of coherent state
for all the time and a coherent Berry phase can be naturally defined
for the coherent state. These facts are proved as follows.

According to the adiabatic theorem proved and the theory of Berry
phase, a system starting from an eigenstate $\psi_{n,0}[\boldsymbol{b}(t=0)]$
will evolve into $e^{-\frac{iE_{n}T}{\hbar}}e^{i\gamma_{n}(T)}\psi_{n,0}[\boldsymbol{b}(T)]$
at time $t=T$ . We note that $E_{n}$ is constant since $|\boldsymbol{B}|=B_{0}$
doesn't change, and the dynamic phase is 
\begin{equation}
\Delta\theta_{d}\equiv\int_{0}^{T}\frac{E_{n}}{\hbar}dt=\frac{E_{n}}{\hbar}T.
\end{equation}
Here, $\gamma_{n}(T)$ is the Berry phase that can be calculated as
\cite{Berry1984} 
\begin{equation}
\gamma_{n}(T)=i\oint_{C}\left<\psi_{n,0}|\frac{\partial}{\partial\boldsymbol{b}}\psi_{n,0}\right>\cdot d\boldsymbol{b}.
\end{equation}
As is calculated in the Appendix, 
\[
\left<\psi_{n,0}|\frac{\partial}{\partial\boldsymbol{b}}\psi_{n,0}\right>\cdot d\boldsymbol{b}.=-in\psi_{n,0}(-d\boldsymbol{e}_{1}\cdot\boldsymbol{e}_{2}).
\]
Therefore, 
\begin{equation}
\gamma_{n}(T)=-n\oint_{C}d\boldsymbol{e}_{1}\cdot\boldsymbol{e}_{2}=n\gamma_{C}(T),\label{eq:Berry}
\end{equation}
where $\gamma_{C}(T)=-\oint_{C}d\boldsymbol{e}_{1}\cdot\boldsymbol{e}_{2}=-\Omega$
is the same as Eq.\,(\ref{eq:solidangle}).

If at $t=0$ the system is at a coherent state $\psi(0)=\Psi_{w_{0}}=e^{-|f|^{2}/2}\sum_{n=0}^{+\infty}(f^{n}/\sqrt{n!})\psi_{n,0}[\boldsymbol{b}(0)]$,
where $f=-i\sqrt{qB_{0}/2\hbar}w_{0}$, then at $t=T$ each eigenstate
component of $\Psi_{w_{0}}$ will gain a Berry phase, and the system
will be 
\begin{align}
\psi(T) & =e^{-\frac{|f|^{2}}{2}}\sum_{n=0}^{+\infty}\frac{f^{n}}{\sqrt{n!}}e^{-\frac{iE_{n}T}{\hbar}}e^{i\gamma_{n}(T)}\psi_{n,0}[\boldsymbol{b}(T)]\nonumber \\
= & e^{-\frac{i\omega_{d}T}{2}}e^{-\frac{|f(T)|^{2}}{2}}\sum_{n=0}^{+\infty}\frac{[f(T)]^{n}}{\sqrt{n!}}\psi_{n,0}[\boldsymbol{b}(0)]\nonumber \\
= & e^{-\frac{i\omega_{d}T}{2}}\Psi_{w_{0}(T)}.\label{eq:psiT}
\end{align}
Here, $f(T)=-i\text{\ensuremath{\sqrt{qB_{0}/2\hbar}}}w_{0}(T)$ and
$w_{0}(T)=w_{0}e^{-i\omega_{d}T}e^{i\gamma_{C}}$. Apparently, the
wave function $\psi(T)$ describes a Gaussian packet centered at $w_{0}(T)$
and of the same size as $\psi(0)$. Thus $\psi(T)$ is still a coherent
state. In $w_{0}(T)$, apart from the dynamic contribution $e^{-i\omega_{d}T}$,
there is also a geometric term $e^{i\gamma_{C}}$ contributing to
the angular position of the wave packet. Therefore, $\gamma_{C}(T)$
can be defined to be the coherent Berry phase of the coherent state,
which is exactly the geometric phase for a classical gyro-motion given
by Eq.\,(\ref{eq:solidangle}). We note that although the Berry phase
is a quantum phase factor, which does not affect the probability distribution
for each eigenstate, the coherent interference of Berry phases $\gamma_{n}(T)$
among all the eigenstates produces an observable effect, which moves
the center of the coherent state by a gyro-phase in the amount of
$\gamma_{C}(T)$ as specified by the phase factor $e^{i\gamma_{C}}$
in $w_{0}(T)$.

\section{Berry phases of a electron with spin}

Liu and Qin \cite{Liu11} compared the geometric phase in the classical
gyro-motion with the Berry phase of the electron spin. But no direct
connection was found. We have shown that the geometric phase of the
gyro-motion is actually the Berry phase associated with the orbital
degree of freedom of a charged particle. To further illustrate the
relationship between these three geometric phases, we solve the Dirac
equation of an electron in this section, and construct, in the non-relativistic
limit, coherent states with spin using the energy eigenstates which
incorporate both the orbital and spin degrees of freedom. This formalism
puts the three geometric phases in one united picture. We will show
that the Berry phase of a coherent state consists of two parts, a
coherent Berry phase due to the orbital motion as discussed in Sec.\,III
and a Berry phase due to the spin. The former is the classical geometric
phase of the gyro-motion, and the latter is a quantum phase factor
with no classical interpretation.

The solution to the Dirac equation of an electron in a uniform magnetic
field can be found in literatures \cite{Kuznetsov2003,Yilmaz2007,Bhattacharya2007}.
Here we rewrite it in a form consistent with the notations in this
paper. For an electron with mass $\mu_{e}$ and charge $q=-e$ in
a magnetic field $\boldsymbol{B}=B_{0}\boldsymbol{e}_{z}$, the Dirac
equation is 
\begin{gather}
i\hbar\frac{\partial\psi}{\partial t}=H\psi,\\
H=c\boldsymbol{\ensuremath{\alpha}}\cdot(\boldsymbol{P}+e\boldsymbol{A})+\beta\mu_{e}c^{2},\\
\boldsymbol{\ensuremath{\alpha}}=\left(\begin{array}{ll}
0~~\boldsymbol{\ensuremath{\sigma}}\\
\boldsymbol{\ensuremath{\sigma}}~~0
\end{array}\right),~\beta=\left(\begin{array}{ll}
I~~~~0\\
0~-I
\end{array}\right),
\end{gather}
where $\psi$ is a 4-component vector and $\boldsymbol{\ensuremath{\sigma}}=(\sigma_{x},\sigma_{y,}\sigma_{z})$
are Pauli matrices. An eigenstate can be written as $\psi=e^{-\frac{iEt}{\hbar}}\left(\varphi,\xi\right)$,
where $\varphi$ and $\xi$ are 2-component vectors. In terms of $\varphi$
and $\xi$ the Dirac equation is 
\begin{flalign}
(E-\mu_{e}c^{2})\varphi=c\boldsymbol{\ensuremath{\sigma}}\cdot(\boldsymbol{P}+e\boldsymbol{A})\xi,\\
(E+\mu_{e}c^{2})\xi=\boldsymbol{c\ensuremath{\sigma}}\cdot(\boldsymbol{P}+e\boldsymbol{A})\varphi,\label{eq:dirac}
\end{flalign}
Eliminating $\xi$ in terms of $\varphi$ gives 
\begin{eqnarray}
(E^{2}-\mu_{e}^{2}c^{4})\varphi & = & c^{2}[\boldsymbol{\ensuremath{\sigma}}\cdot(\boldsymbol{P}+e\boldsymbol{A})]^{2}\varphi.
\end{eqnarray}
Using the kinetic momentum operator $\boldsymbol{\ensuremath{\pi}}=\boldsymbol{P}+e\boldsymbol{A}$,
and ignoring the parallel motion $\pi_{z}$, we have 
\begin{flalign}
 & [\boldsymbol{\ensuremath{\sigma}}\cdot(\boldsymbol{P}+e\boldsymbol{A})]^{2}=\left(\begin{array}{ll}
0~~~~~~~\pi_{x}-i\pi_{y}\\
\pi_{x}+i\pi_{y}~~~~~~~0
\end{array}\right)^{2}\nonumber \\
 & =\left(\begin{array}{ll}
\pi_{x}^{2}+\pi_{y}^{2}+i[\pi_{x},\pi_{y}] & \ \ \ \ \ \ \ \ \ \ \ \ \ \ \ \ \ 0\\
0 & \pi_{x}^{2}+\pi_{y}^{2}-i[\pi_{x},\pi_{y}]
\end{array}\right),
\end{flalign}
which is diagonalized. Because $[\pi_{x},\pi_{y}]=-i\hbar eB_{0}$
due to the negative electron charge $q=-e$, we redefine the creating
and annihilating operators as 
\begin{gather}
a^{\dagger}=\sqrt{\frac{1}{2\hbar eB_{0}}}(\pi_{y}-i\pi_{x}),\\
a=\sqrt{\frac{1}{2\hbar eB_{0}}}(\pi_{y}+i\pi_{x}),
\end{gather}
so that $\omega_{d}$ defined below can be positive. Then, 
\begin{equation}
[\boldsymbol{\ensuremath{\sigma}}\cdot(\boldsymbol{P}+e\boldsymbol{A})]^{2}=2\mu_{e}\hbar\omega_{d}\left(\begin{array}{ll}
a^{\dagger}a+1~~~~~0\\
0~~~~~~~~~~~a^{\dagger}a
\end{array}\right),
\end{equation}
where $\omega_{d}=eB_{0}/\mu_{e}>0$ . Let $\varphi=\left(\varphi_{+},\varphi_{-}\right)$
and express the Dirac equation for $\varphi$ as 
\begin{eqnarray}
(E^{2}-\mu_{e}^{2}c^{4})\varphi_{+} & = & 2\mu_{e}c^{2}\hbar\omega_{d}(a^{\dagger}a+1)\varphi_{+},\\
(E^{2}-\mu_{e}^{2}c^{4})\varphi_{-} & = & 2\mu_{e}c^{2}\hbar\omega_{d}(a^{\dagger}a)\varphi_{-}.
\end{eqnarray}
The eigenstates $\psi_{n,m}$ on Landau levels can be obtained using
the same procedure in Sec.\,III: 
\begin{flalign}
\varphi_{+}=\psi_{n,m}, & \ E_{+}=\sqrt{\mu_{e}^{2}c^{4}+2(n+1)\hbar\omega_{d}\cdot\mu_{e}c^{2}},\\
\varphi_{-}=\psi_{n',m'}, & \ E_{-}=\sqrt{\mu_{e}^{2}c^{4}+2n'\hbar\omega_{d}\cdot\mu_{e}c^{2}}.
\end{flalign}
The Landau levels are relativistic, and there is a difference between
$E_{+}$ and $E_{-}$ due to the spin. Here, $\varphi_{+}$ and $\varphi_{-}$
are required to have the same energy, i.e., $E_{+}=E_{-}=E$, but
we can let one of them to be zero and obtain a set of solutions as
$\left(\psi_{n,m},0\right)$ and $\left(0,\psi_{n,m}\right)$. Once
the $\varphi$ component is known, the $\xi$ component can be calculated
directly from Eq.\,(\ref{eq:dirac}). In the non-relativistic limit,
$E\approx\mu_{e}c^{2}$ and $\xi$ is negligible compared to $\varphi$.
Hence, a set of solutions to the Dirac equation in the non-relativistic
limit is 
\begin{equation}
\psi_{+,n,m}=\left(\begin{array}{ll}
\psi_{n,m}\\
0\\
0\\
0
\end{array}\right),~\psi_{-,n,m}=\left(\begin{array}{ll}
0\\
\psi_{n,m}\\
0\\
0
\end{array}\right),\label{eq:landau_spin}
\end{equation}
which incorporates both the orbital and spin degrees of freedom. During
the adiabatic variation of the magnetic field $\boldsymbol{B}(t)=B_{0}\boldsymbol{b}(t)$,
each eigenstate will become the instantaneous eigenstate. Since the
magnetic field only changes its direction, the instantaneous eigenstates
for the Hamiltonian $H[\boldsymbol{b}(t)]$ can be obtained by applying
a Lorentz transformation to the eigenstates specified by Eq.\,(\ref{eq:landau_spin})
\cite{QFT}, 
\begin{equation}
\psi(x)\rightarrow\Lambda_{\frac{1}{2}}\psi(L^{-1}x),
\end{equation}
where $L$ is a spatial transformation which rotates $\boldsymbol{e}_{z}$
to $\boldsymbol{b}$. If we use spherical coordinates $\boldsymbol{b}=(\sin\zeta\cos\phi,\sin\zeta\sin\phi,\cos\zeta)$,
then $L$ can be a rotation around the axis passing through the origin
and in the direction of $\boldsymbol{\ensuremath{\omega}}=(-\zeta\sin\phi,\zeta\cos\phi,0)$,
where $|\boldsymbol{\ensuremath{\omega}}|=\zeta$ is the rotation
angle. The corresponding transformation on the spin components is
$\Lambda_{\frac{1}{2}}=e^{-i\boldsymbol{\ensuremath{\omega}}\cdot\boldsymbol{S}/2}$
, where $\boldsymbol{S}=\left(\begin{array}{ll}
\boldsymbol{\ensuremath{\sigma}}~~0\\
0~~\boldsymbol{\ensuremath{\sigma}}
\end{array}\right)$. A simple calculation shows that 
\begin{equation}
\Lambda_{\frac{1}{2}}=\left(\begin{array}{ll}
S_{\frac{1}{2}} & \ 0\\
0 & S_{\frac{1}{2}}
\end{array}\right),\ S_{\frac{1}{2}}=\left(\begin{array}{ll}
\cos\frac{\zeta}{2}\ \ \ -e^{-i\phi}\sin\frac{\zeta}{2}\\
e^{i\phi}\sin\frac{\zeta}{2}\ \ \ \ \ \ \cos\frac{\zeta}{2}
\end{array}\right).
\end{equation}
Thus the instantaneous eigenstates on the Landau levels of the Hamiltonian
$H[\boldsymbol{b}(t)]$ are 
\begin{equation}
\psi_{\pm,n,m}=\psi_{n,m}[\boldsymbol{b}(t)]\left|\pm\right>[\boldsymbol{b}(t)],
\end{equation}
where $\psi_{n,m}[\boldsymbol{b}(t)]$ is the same as that in Sec.
IV, and $\left|\pm\right>[\boldsymbol{b}(t)]$ are instantaneous spin
eigenstates, 
\begin{equation}
\left|+\right>[\boldsymbol{b}]=\left(\begin{array}{ll}
\cos\frac{\zeta}{2}\\
e^{i\phi}\sin\frac{\zeta}{2}\\
0\\
0
\end{array}\right),\ \left|-\right>[\boldsymbol{b}]=\left(\begin{array}{ll}
-e^{-i\phi}\sin\frac{\zeta}{2}\\
\cos\frac{\zeta}{2}\\
0\\
0
\end{array}\right).
\end{equation}
When $\boldsymbol{b}(t)$ varies slowly with time, we can calculate
the Berry phase for each $\psi_{\pm,n,0}$ as follows, 
\begin{align}
\gamma_{\pm,n}(T)= & i\oint\left<\psi_{\pm,n,0}|\frac{\partial}{\partial\boldsymbol{b}}|\psi_{\pm,n,0}\right>\nonumber \\
= & i\oint\left<\psi_{n,0}|\frac{\partial}{\partial\boldsymbol{b}}|\psi_{n,0}\right>+i\oint\left<\pm|\frac{\partial}{\partial\boldsymbol{b}}|\pm\right>=(n\pm\frac{1}{2})\gamma_{C}(T),\label{eq:gamma}
\end{align}
where $\gamma_{C}=\oint d\boldsymbol{e}_{1}\cdot\boldsymbol{e}_{2}=\Omega$,
$n\gamma_{C}(T)$ is the orbital Berry phase and $\pm\gamma_{C}(T)/2$
is the spin Berry phase. Here the sign of $\gamma_{C}(T)$ is different
from that in Eq.\,(\ref{eq:Berry}) due to the negative electron
charge.

We can construct spin-up coherent states using $\psi_{+,n,0}$ or
spin-down coherent states using $\psi_{-,n,0}$ as in Sec.\,III,
\begin{flalign}
\Psi_{\pm,w_{0}}=e^{-\frac{|f|^{2}}{2}}e^{fa^{\dagger}}\psi_{\pm,0,0}=e^{-\frac{|f|^{2}}{2}}\sum_{n=0}^{+\infty}\frac{f^{n}}{\sqrt{n!}}\psi_{\pm,n,0} & ,\label{eq:coherent0-1}
\end{flalign}
where $f=-i\sqrt{qB_{0}/2\hbar}w_{0}$. Note that since the definition
for $a^{\dagger}$ has changed due to the negative electron charge,
the coherent states defined by Eq.\,(\ref{eq:coherent0-1}) are actually
centered in $\bar{w_{0}}$. The evolution of $\Psi_{\pm,w_{0}}$ when
$\boldsymbol{b}(t)$ slowly varies follows the same derivation of
Eq.\,(\eqref{eq:psiT}): 
\begin{align}
\Psi_{\pm,w_{0}}(T) & =e^{-\frac{|f|^{2}}{2}}\sum_{n=0}^{+\infty}\frac{f^{n}}{\sqrt{n!}}e^{-\frac{iE_{n}T}{\hbar}}e^{i\gamma_{n}(T)}\psi_{\pm,n,0}[\boldsymbol{b}(T)]\nonumber \\
= & e^{-\frac{i\omega_{d}T}{2}}e^{\pm\frac{i\gamma_{C}(T)}{2}}e^{-\frac{|f(T)|^{2}}{2}}\sum_{n=0}^{+\infty}\frac{[f(T)]^{n}}{\sqrt{n!}}\psi_{\pm,n,0}[\boldsymbol{b}(0)]\nonumber \\
= & e^{-\frac{i\omega_{d}T}{2}}e^{\pm\frac{i\gamma_{C}(T)}{2}}\Psi_{\pm,w_{0}(T)},\label{eq:psiT-1}
\end{align}
where $f(T)=-i\sqrt{qB_{0}/2\hbar}w_{0}(T)$ and $w_{0}(T)=w_{0}e^{-i\omega_{d}T}e^{i\gamma_{C}}$.
It is clear that $\Psi_{\pm,w_{0}}(T)$ are still coherent states
with spin-up or spin-down. As in the case without spin, the orbital
Berry phases for each $\psi_{\pm,n,0}$ interfere coherently to produce
a coherent Berry phase corresponding to the classical geometric phase
of the gyro-motion. The Berry phases for the spin degree of freedom
remain to be quantum phase factors for the coherent states, bringing
no classical effect.

\section{Conclusions }

In this paper, we have shown the correspondence between the geometric
phase of the classical gyro-motion of a charged particle in a slowly
varying magnetic field and the quantum Berry phase of the orbital
degree of freedom. This task is accomplished by first constructing
a coherent state for a spinless particle using the energy eigenstates
on the Landau levels and proving that the coherent states can maintain
their status of coherent states during the adiabatic varying of the
magnetic field. It is discovered that for the coherent state, a coherent
Berry phase can be naturally defined, which is exactly the classical
geometric phase of the gyro-motion.

To include the spin dynamics into the analysis, we have also studied
electrons with spin described by the Dirac equation. Using the energy
eigenstates which incorporate both the orbital and spin degrees of
freedom, we have shown that in the non-relativistic limit, spin-up
or spin-down coherent states can be constructed. For each of the eigenstate
that makes up the coherent states, the Berry phase consists of two
parts that can be identified as those due to the orbital and spin
motion. For the coherent states, the orbital Berry phases of eigenstates
interfere coherently such that a coherent Berry phase can be naturally
defined, which is exactly the geometric phase of the classical gyro-motion.
The spin Berry phases of the eigenstates, on the other hand, remains
to be quantum phase factors for the coherent state and have no classical
counterpart.

There are interesting topics worthy of further investigation. The
first is that it is not obvious that a classical particle must be
represented by a non-diffusive Gaussian wave packet. Any wave packet
that is localized and evolves stably with time can be a candidate.
For example, other ground states $\psi_{0,m}$ ($m>0$) can also be
used to generate coherent states, and indeed we find that $\Psi_{w_{0}}=e^{-|f|^{2}/2}e^{fa^{\dagger}}\psi_{0,m}$
are also coherent states with more complicated structures. There is
also a way of constructing a coherent state whose wave packet does
not even have rotational symmetry around its center \cite{Kowalsk2005}.
For these constructions, we should be able to establish the connection
between quantum Berry phases and classical geometric phases using
the same techniques developed here. Another related topic is that
the spatial non-uniformity of magnetic field can also give geometric
phases \cite{Littlejohn1988,Brizard2012,Liu12-2}. A quantum treatment
for gradient-B drift has been developed \cite{Chan16-1,Chan16-2}.
However, the construction of coherent states in inhomogeneous magnetic
field requires more sophisticated techniques which are beyond the
scope of this paper and will be discussed elsewhere. 
\begin{acknowledgments}
We thank Junyi Zhang and Jian Liu for fruitful discussion. This research
was supported by the U.S. Department of Energy (DE-AC02-09CH11466). 
\end{acknowledgments}

\section*{appendix: proof to the adiabatic theorem for the eigenstates on the
landau levels }

Here we give a proof to to the adiabatic theorem for eigenstates $\psi_{n,m}$
on the Landau levels. Specifically, we prove that when the magnetic
field changes its direction very slowly, i.e., $\boldsymbol{B}(t)=B_{0}\boldsymbol{b}(t)$
for a slowly varying $\boldsymbol{b}(t)$, each energy eigenstate
$\psi_{n,m}[\boldsymbol{b}(0)]$ at $t=0$ will evolve independently,
and at later time $t$ will be on the energy eigenstate $\psi_{n,m}[\boldsymbol{\boldsymbol{b}}(t)]$
determined by $\boldsymbol{b}(t)$ at the instant of $t.$ In section.\,III-V,
the motion $\boldsymbol{P}_{z}$ along the magnetic field was ignored
since the parallel motion is decoupled from the perpendicular motion,
and the eigenstates on Landau levels are invariant under parallel
translation. However, this translational symmetry breaks down when
$\boldsymbol{B}$ changes its direction, thus we must consider the
parallel motion in this proof.

To evaluate transition amplitudes, integration of the wave functions
along $\boldsymbol{B}$ are needed. For this purpose, we consider
a system which has finite extension, i.e., $-L/2<z[\boldsymbol{b}]<L/2$,
where 
\begin{equation}
z[\boldsymbol{b}]=\boldsymbol{r}\cdot\boldsymbol{b}\label{eq:polarz}
\end{equation}
is the distance along $\boldsymbol{b}$ in cylindrical coordinate.
In general, we can choose $L$ to be one or two orders larger than
the transverse dimension of the wave function. We also assume periodic
boundary conditions in the $z$-direction. Normalized eigenstate wave
functions and energies of the Schrödinger equation for a charged particle
in a uniform magnetic field can be easily obtained: 
\begin{align}
\psi_{n,m,l} & =\frac{1}{\sqrt{L}}\psi_{n,m}(\rho[\boldsymbol{b}],\theta[\boldsymbol{b}])e^{i\frac{2\pi l}{L}z[\boldsymbol{b}]},\label{eigenstates}\\
E_{n,l} & =\hbar\omega_{d}(n+\frac{1}{2})+\frac{2\pi^{2}\hbar^{2}l^{2}}{\mu L^{2}},
\end{align}
where $\rho[\boldsymbol{b}]$ and $\phi[\boldsymbol{b}]$ are defined
in Eqs.\,(\ref{eq:polarrho}) and (\ref{eq:polartheta}). $\psi_{n,m}(\rho,\theta)=R_{n,m}(\rho)e^{-i(n-m)\theta}$
are eigenstates on the Landau levels, and $R_{n,m}(\rho)$ are real.
The quantized parallel motion are labeled by $l=0,\pm1,\pm2,...$,
corresponding to the momentum $p_{z}=2\pi\hbar l/L$. The Schrödinger
equation with a time-dependent Hamiltonian $H[\boldsymbol{b}(t)]$
is 
\begin{equation}
i\hbar\frac{\partial}{\partial t}\left|\psi(t)\right>=H[\boldsymbol{b}(t)]\left|\psi(t)\right>.
\end{equation}
In general, $\left|\psi(t)\right>$ is the superposition of all the
eigenstates of $H[\boldsymbol{b}(t)]$, 
\begin{equation}
\left|\psi(t)\right>=\sum_{n,m,l}a_{n,m,l}(t)e^{-\frac{iE_{n,l}t}{\hbar}}\left|\psi_{n,m,l}[\boldsymbol{b}(t)]\right>.
\end{equation}
Inserting this expression into the Schrödinger equation, and taking
the inner product with $\left<\psi_{n,m,l}[\boldsymbol{b}(t)]\right|$,
we obtain the dynamic equation for the coefficients $a_{n,m,l}(t)$,
\begin{align}
\frac{d}{dt}a_{n,m,l}(t) & =-a_{n,m,l}(t)\left<\psi_{n,m,l}[\boldsymbol{b}(t)]|\frac{\partial}{\partial t}\psi_{n,m,l}[\boldsymbol{b}(t)]\right>\nonumber \\
-\sum_{(n',m',l')} & a_{n',m',l'}(t)e^{-\frac{i(E_{n,l}-E_{n',l'})t}{\hbar}}\left<\psi_{n,m,l}[\boldsymbol{b}(t)]|\frac{\partial}{\partial t}\psi_{n',m',l'}[\boldsymbol{\boldsymbol{b}}(t)]\right>,\label{eq:da}
\end{align}
where the summation is over all the $(n',m',l')\neq(n,m,l)$. The
adiabatic theorem states that after integrating over time, the contribution
from the summation term can be neglected if \cite{Messiah1962,Holstein1989}
\begin{equation}
\left|\frac{\hbar\left<\psi_{n,m,l}|\frac{\partial}{\partial t}\psi_{n',m',l'}\right>}{E_{n,l}-E_{n',l'}}\right|\sim\epsilon\ll1,~\forall(n',m',l')\neq(n,m,l),\label{eq:adiabatic2}
\end{equation}
then each $a_{n,m,l}$ evolves separately, and we are able to conclude
that each eigenstate remains to be the instantaneous eigenstate. However,
if $E_{n,l}-E_{n',l'}=0$, then condition (\ref{eq:adiabatic2}) cannot
be satisfied unless $\left<\psi_{n,m,l}|\frac{\partial}{\partial t}\psi_{n',m',l'}\right>$
is strictly $0$. Here we prove that if $l=0$, then $\left<\psi_{n,m,l=0}|\frac{\partial}{\partial t}\psi_{n',m',l'}\right>$
is indeed $0$ when $E_{n,l=0}-E_{n',l'}=0$. Thus the adiabatic theorem
is valid on Landau levels, particular for $\psi_{n,0}$ which makes
up the coherent state $\Psi_{w_{0}}$ in Eq.\,(\ref{eq:coherent0}).
There are two possible situations when $E_{n,l=0}-E_{n',l'}$ could
be $0$. The first is when $n'=n$, $m'\neq m$, $l'=0$, which can
always happen. The second is when $n'<n$, $2\pi^{2}\hbar^{2}l'^{2}/\mu L^{2}=(n-n')\hbar\omega_{d}$,
i.e., the energy from parallel motion fills the gap between two Landau
levels. The second situation only happens if $\sqrt{\mu\omega_{d}L^{2}/2\pi^{2}\hbar}$
is an integer, and can be avoided by choosing an $L$ such that $\sqrt{\mu\omega_{d}L^{2}/2\pi^{2}\hbar}$
is not an integer.

Let's prove that for the first situation ($n'=n$, $m'\neq m$, $l'=0$),
$\left<\psi_{n,m,0}|\frac{\partial}{\partial t}\psi_{n',m',l'}\right>$
is always 0. By the chain rule, the time derivative is

\begin{flalign}
 & \frac{\partial}{\partial t}\psi_{n',m',l'}\{\rho[\boldsymbol{b}(t)],\theta[\boldsymbol{b}(t)],z[\boldsymbol{b}(t)]\}=\nonumber \\
 & \left[\frac{\partial\psi_{n',m',l'}}{\partial\rho}\frac{\partial\rho[\boldsymbol{b}]}{\partial\boldsymbol{b}}+\frac{\partial\psi_{n',m',l'}}{\partial\theta}\frac{\partial\theta[\boldsymbol{b}]}{\partial\boldsymbol{b}}+\frac{\partial\psi_{n',m',l'}}{\partial z}\frac{\partial z[\boldsymbol{b}]}{\partial\boldsymbol{b}}\right]\cdot\frac{d\boldsymbol{b}(t)}{dt},\label{eq:ppsi}
\end{flalign}
and from the definitions of $\rho[\boldsymbol{b}]),\ \theta[\boldsymbol{b}],\ z[\boldsymbol{b}]$
in Eqs.\,(\ref{eq:polarrho}), (\ref{eq:polartheta}), and (\ref{eq:polarz}),
we have 
\begin{align}
\frac{\partial\rho[\boldsymbol{b}]}{\partial\boldsymbol{b}}\cdot\frac{d\boldsymbol{b}}{dt} & =-\frac{(\boldsymbol{r}\cdot\boldsymbol{b})}{\rho}\boldsymbol{r}\cdot\frac{d\boldsymbol{b}}{dt}=-\frac{z}{\rho}\boldsymbol{\rho}\cdot\frac{d\boldsymbol{b}}{dt},\\
\frac{\partial\theta[\boldsymbol{b}]}{\partial\boldsymbol{b}}\cdot\frac{d\boldsymbol{b}}{dt} & =-\boldsymbol{\ensuremath{\boldsymbol{e_{2}}}}\cdot\frac{d\boldsymbol{\ensuremath{\boldsymbol{e_{1}}}}}{dt}-\frac{z}{\rho}\cos\theta\boldsymbol{\ensuremath{\boldsymbol{e_{2}}}}\cdot\frac{d\boldsymbol{b}}{dt}+\frac{z}{\rho}\sin\theta\boldsymbol{\ensuremath{\boldsymbol{e_{1}}}}\cdot\frac{d\boldsymbol{b}}{dt},\label{eq:partialtheta}\\
\frac{\partial z[\boldsymbol{b}]}{\partial\boldsymbol{b}}\cdot\frac{d\boldsymbol{b}}{dt} & =\boldsymbol{r}\cdot\frac{d\boldsymbol{b}}{dt}=\boldsymbol{\rho}\cdot\frac{d\boldsymbol{b}}{dt},
\end{align}
where $\boldsymbol{\rho}=\cos\theta\boldsymbol{e}_{1}+\sin\theta\boldsymbol{e}_{2}$.
Let $d\boldsymbol{b}=\cos\theta_{0}\boldsymbol{e}_{1}+\sin\theta_{0}\boldsymbol{e}_{2}$
(since $d\boldsymbol{b}\cdot\boldsymbol{b}=0$), then $\boldsymbol{\rho}\cdot d\boldsymbol{b}=\rho|d\boldsymbol{b}|\cos(\theta-\theta_{0})$.
Note that $\boldsymbol{e}_{2}\cdot d\boldsymbol{e}_{1}/dt$, $\boldsymbol{e}_{2}\cdot d\boldsymbol{b}/dt$
and $\boldsymbol{e}_{1}\cdot d\boldsymbol{b}/dt$ are constant for
the spatial integration. Putting these results into Eq.\,(\ref{eq:ppsi}),
we have 
\begin{align}
\left<\psi_{n,m,0}|\frac{\partial}{\partial t}\psi_{n',m',l'}\right> & =\int_{0}^{+\infty}\rho d\rho\int_{0}^{2\pi}d\theta\int_{-\frac{L}{2}}^{\frac{L}{2}}dz\psi_{n,m,0}^{*}\times\nonumber \\
 & \left[\frac{\partial\psi_{n',m',l'}}{\partial\rho}\frac{\partial\rho}{\partial\boldsymbol{b}}+\frac{\partial\psi_{n',m',l'}}{\partial\theta}\frac{\partial\theta}{\partial\boldsymbol{b}}+\frac{\partial\psi_{n',m',l'}}{\partial z}\frac{\partial z}{\partial\boldsymbol{b}}\right]\cdot\frac{d\boldsymbol{b}}{dt}.\label{eq:integration}
\end{align}
For the eigenstate wave functions $\psi_{n.m,l}=R_{n,m}(\rho)e^{-i(n-m)\theta}e^{i\frac{2\pi l}{L}z}/\sqrt{L}$,
the first integration in Eq.\,(\ref{eq:integration}) is 
\begin{align}
 & \left<\psi_{n,m,0}|\frac{\partial\psi_{n',m',l'}}{\partial\rho}\frac{\partial\rho}{\partial\boldsymbol{b}}\cdot\frac{d\boldsymbol{b}}{dt}\right>=-|\frac{d\boldsymbol{b}}{dt}|\frac{1}{L}\int_{-\frac{L}{2}}^{\frac{L}{2}}ze^{i\frac{2\pi l'}{L}z}dz\times\nonumber \\
 & \int_{0}^{+\infty}\int_{0}^{2\pi}\rho d\rho d\theta R_{n,m}(\rho)\frac{dR_{n',m'}(\rho)}{d\rho}e^{i(n-m)\theta}e^{-i(n'-m')\theta}\cos(\theta-\theta_{0}).\label{eq:interho}
\end{align}
We see that the $z$ integration is $0$ if $l'=0$. The second integration
in Eq.\,(\ref{eq:integration}) has three terms due to the expression
of $\frac{\partial\theta[\boldsymbol{b}]}{\partial\boldsymbol{b}}\cdot\frac{d\boldsymbol{b}}{dt}$
from Eq.\,(\ref{eq:partialtheta}). The integration containing $\boldsymbol{e}_{2}\cdot d\boldsymbol{e}_{1}/dt$
is strictly zero because $\frac{\partial}{\partial\theta}\psi_{n',m',l'}=-i(n'-m')\psi_{n',m',l'}$
and $\left<\psi_{n,m,0}|\psi_{n',m',l'}\right>=0$ if $m'\neq m$.
The integration containing $-(z/\rho)\cos\theta\boldsymbol{e}_{2}\cdot d\boldsymbol{b}/dt$
is 
\begin{align}
 & -i(n'-m')(-\boldsymbol{e}_{2}\cdot\frac{d\boldsymbol{b}}{dt})\left<\psi_{n,m,0}|\frac{z}{\rho}\cos\theta\psi_{n',m',l'}\right>=-|\frac{d\boldsymbol{b}}{dt}|\frac{1}{L}\int_{-\frac{L}{2}}^{\frac{L}{2}}ze^{i\frac{2\pi l'}{L}z}dz\times\nonumber \\
 & \int_{0}^{+\infty}\int_{0}^{2\pi}d\rho d\theta R_{n,m}(\rho)R_{n',m'}(\rho)e^{i(n-m)\theta}e^{-i(n'-m')\theta}\cos(\theta-\theta_{0}),\label{eq:intetheta}
\end{align}
in which the $z$ integration also gives $0$ if $l'=0$. The integration
containing $(z/\rho)\sin\theta\boldsymbol{e}_{1}\cdot d\boldsymbol{b}/dt$
can be calculated in the same way. Finally, the third integration
in Eq.\,(\ref{eq:integration}) is 
\begin{align}
 & \left<\psi_{n,m,0}|\frac{\partial\psi_{n',m',l'}}{\partial z}\frac{\partial z}{\partial\boldsymbol{b}}\cdot\frac{d\boldsymbol{b}}{dt}\right>=-|\frac{d\boldsymbol{b}}{dt}|\frac{1}{L}\int_{-\frac{L}{2}}^{\frac{L}{2}}(i\frac{2\pi l'}{L})e^{i\frac{2\pi l'}{L}z}dz\times\\
 & \int_{0}^{+\infty}\int_{0}^{2\pi}\rho d\rho d\theta R_{n,m}(\rho)R_{n',m'}(\rho)e^{i(n-m)\theta}e^{-i(n'-m')\theta}\cdot\rho\cos(\theta-\theta_{0}),\label{eq:intez}
\end{align}
which is always $0$ due to the integration in $z$. Thus, we proved
that $\left<\psi_{n,m,0}|\frac{\partial}{\partial t}\psi_{n',m',l'}\right>=0$
when $n'=n$, $m'\neq m$, $l'=0$, and therefore proved the adiabatic
theorem for eigenstates $\psi_{n,m}$ on Landau levels.

\end{document}